\documentstyle[12pt,psfig,aaspp4]{article}
\newcommand{\beq}{\begin{equation}}
\newcommand{\eeq}{\end{equation}}
\def\_#1{_{\scriptscriptstyle #1}}
\def\&#1{^{\scriptscriptstyle #1}}
\def\rar{\rightarrow}
\def\rg{{\rm g}}
\def\a0{a_0}
\def\gh{\rg_h}
\def\vgh{{\bf g}_h}
\def\abs#1{\vert #1\vert}
\def\gn{\rg\_{N}}
\def\vgn{{\bf g}\_{N}}
\def\vg{{\bf g}}
\def\gx{\rg_{max}}
\def\div{{\vec\nabla}\cdot}
\def\grad{{\vec\nabla}}

\def\f{\varphi}
\def\gf{\grad\f}
\begin{document}

\title{The modified dynamics (MOND) predicts an absolute maximum
to the acceleration produced by `dark halos'}
\author{ Rafael Brada and Mordehai Milgrom}
\affil{Department of Condensed Matter Physics, Weizmann Institute of
Science, Rehovot 76100 Israel}

\begin{abstract}
We have recently discovered that the modified dynamics (MOND)
implies some universal upper bound on
the acceleration that can be contributed by a `dark halo'--assumed 
in a Newtonian analysis to account for the effects of MOND.
Not surprisingly, the limit is of the order of the acceleration
constant of the
theory. This can be contrasted directly with the results of
structure-formation simulations. The new limit is substantial and different
from earlier MOND acceleration limits (discussed in connection with the MOND
explanation of the Freeman
law for galaxy disks, and the Fish law for ellipticals):
It pertains to the `halo', and not to the observed galaxy;
it is absolute, and independent of further 
physical assumptions on the nature of the galactic system; and
it applies at all radii, whereas the other limits apply only to the 
mean acceleration in the system.

\end{abstract}
\keywords{gravitation-galaxies: halos, kinematics and dynamics}

\section {Introduction}
\label{introduction}
\par
The acceleration constant of the modified dynamics (MOND), $\a0$, appears 
in various predicted regularities pertinent to galaxies.
For example, it features as an upper cutoff to the mean surface density
(or mean surface brightness--translated with $M/L$) of galaxies,
 as observed and formulated in
 the Freeman law for disks, and of the Fish law for ellipticals.
We have now come across another such role of $\a0$
that had escaped our notice until
recently: In spherical configurations, and in those relevant to
 rotation-curve analysis of disk galaxies,
the excess, $\gh\equiv \rg-\gn$, of the MOND acceleration, $\rg$, 
over the Newtonian value
for the same mass, $\gn$,
is universally bounded from above by a value $\gx=\eta\a0$, where $\eta$
is of order 1.
Thus, if we attribute what are the effects of MOND to
the presence of a fictitious
dark halo, $\gx$ is a
 universal upper bound to the acceleration produced by the `halo', in all
systems, and at all radii.
If the `halo' is assumed quasi-spherical, this can be put as a statement on
the accumulated (three dimensional) surface density of the `halo',
 which must obey the 
universal bound $M_h(r)/r^2\le \eta\a0 G^{-1}$.
\par
Inasmuch as MOND is successful in explaining the rotation curves of disk
galaxies with reasonable stellar $M/L$ values (\cite{san,sanver,demc})
 we can deduce that, indeed, `halo' accelerations are bounded by $\gx$.
 This is an important observation
regardless of whether MOND entails new physics, or is just an economical
 way of describing dark halos.
Newtonian, disk-plus-dark-halo decompositions and rotation-curve fits
 are rather more
flexible because they involve two added parameters for the halo, allowing
 one to maximize the contribution of the halo,
minimizing that of the disk. But, reasonable fits do give 
a maximum halo acceleration. For example, Sanders (private communication) 
finds in the dark-halo best fits of \cite{bbs} a maximum acceleration
of $\sim 0.4\a0$ for all the galaxies with reasonable fits. 
\par
We derive this upper bound and explain the assumptions that go into the
derivation in section 2.
Then, in section 3, we compare this new limit with previous MOND limits
on the acceleration in galactic systems.

\section{derivation of the upper bound}
\label {derivation}
\par
The absolute upper bound on $\gh$ follows simply from the basic MOND
relation between the acceleration $\rg$ and the Newtonian acceleration
$\gn$:
\beq \mu(\rg/\a0)\rg=\gn,   \label{basica} \eeq
$\mu(x)$ being the interpolating function of MOND. The validity of
 this relation constitutes part of the underlying assumptions (see below).
The excess acceleration $\gh=\rg-\gn$ can be written as a function of $\rg$:
\beq \gh=\rg-\rg\mu(\rg/\a0).  \label{giga} \eeq
Now, $\rg$ can take any (non-negative) value, but, for all
 acceptable forms of 
$\mu(x)$, expression (\ref{giga}) has a maximum, which $\gh$ can thus not 
exceed. Writing $x=\rg/\a0$, and $y=\gh/\a0$,  $y(x)=x[1-\mu(x)]$ is 
non-negative and vanishes at $x=0$. Thus, it has a global
 maximum if and only if 
it does not diverge at $x\rar\infty$; i.e., if $\mu(x)$ approaches 1 at
$x\rar\infty$ (as it must do) no slower than $x^{-1}$.
The parameter $\eta$ defined above is just this maximum value of $y(x)$.
There are solar-system constraints on how slowly $\mu(x)$ can approach 1
in the Newtonian limit (\cite{mond}). Such constraints practically 
exclude the possibility that $y(x)$ diverges at large $x$.  
Some examples: for $\mu(x)=x/(1+x)$ the maximum, achieved in the Newtonian
 limit, is $\eta=1$; for the often-used $\mu(x)=x(1+x^2)^{-1/2}$,
 $\eta= [(\sqrt{5}-1)/2]^{5/2}\approx 0.3$; for $\mu(x)=1-e^{-x}$,
 $\eta=e^{-1}\approx 0.37$. (We see that, in fact, $\eta$ tends to be
 rather smaller than 1.)
\par
When is expression (\ref{basica}) valid?
MOND may be viewed as either a modification of gravity or as one of inertia.
Mondified gravity is described by the generalized Poisson equation discussed
in \cite{bm}, which is of the form
\beq \div[\mu(\abs{\gf}/\a0)\gf]=4\pi G\rho, \label{ppp}\eeq
where $\f$ is the (MOND) potential produced by the mass distribution $\rho$.
 For systems with one-dimensional symmetry (e.g. in spherically
symmetric ones) eq.(\ref{basica}) is exact in this theory. It was also shown
to be a good approximation for the acceleration in the mid-plane of disk
galaxies (\cite{sol},\cite{brad}). An exact statement that can be made in this
case for an arbitrary mass configuration is that the average value of
 $\abs{\vgh}$ over an equipotential surface
of the `halo' is bounded by $\gx$. To see this note that from eq.(\ref{ppp})
\beq \div\vgh=\div[\vg-\mu(\rg/\a0)\vg]  \label{opluta} \eeq
(because $\div\vgn=4\pi G\rho=\div[\mu(\rg/\a0)\vg]$). Take a Gauss integral
for a volume bounded by an equipotential of $\f_h\equiv\f-\f_N$. Because
$\vgh$ is perpendicular to the surface we have
\beq \int[1-\mu(\rg/\a0)]\vg\cdot {\bf ds}=\int\vgh\cdot{\bf ds}
=\int\abs{\vgh}ds. \label{jopaga} \eeq
Since we proved that $[1-\mu(\rg/\a0)]\rg\le \gx$,
the left-hand side is bounded by $\gx\int ds$, and so
$\langle\abs{\vgh}\rangle\equiv \int\abs{\vgh}ds/\int ds\le\gx$.
\par
There is no concrete theory of mondified inertia yet; but, as was shown in
\cite{ann}, eq.(\ref{basica}) is exact in all such theories
 for circular orbits
in an axisymmetric potential. So our limit here 
would apply, in both versions of MOND,
 to the `halo' deduced from rotation-curve analysis.
\section{comparison with previous MOND acceleration limits}
\label {comparison}
The acceleration constant of MOND, $\a0$, has been found before to
define a sort of limiting acceleration in two cases.
The first case concerns self-gravitating spheres supported by random motions
with constant tangential and radial velocity dispersions. 
The mean acceleration in all such spheres cannot exceed a certain value
of order $\a0$ (\cite{is}). This was suggested as an explanation of
 the Fish law, by which
the distribution of the central surface brightnesses in ellipticals is 
sharply cutoff above a certain value (which, assuming some typical $M/L$
 value, translates into a 
  mean surface density $\Sigma \sim\a0 G^{-1}$).
 The second instance concerns
 self-gravitating disks. In MOND, disks with a
mean acceleration much larger than $\a0$ are in the
 Newtonian regime and are less stable than disks in the MOND regime,
 with mean
accelerations smaller than $\a0$ (\cite{stability,bmstab} and references
 therein). This was
 suggested as an explanation of the Freeman law in its revised form,
whereby the distribution of central surface brightnesses of galactic disks
is cut off above a certain value (see a recent review and further references 
in \cite{mccut}).
\par
The new limit we discuss here is different from those two
 in several important regards.
\par
1.  The previous limits concern the visible part of the galaxy, while the
new limit pertains to the fictitious halo and thus lends itself to direct
comparison with predictions of structure-formation simulations, which
are rather vague as regards the visible galaxy.
 At the moment such simulations are also equivocal on the exact
 structure of the halo itself. Different simulations start with different 
assumptions, and the effect of the visible galaxy on the halo is also
poorly accounted for. Nonetheless, it may be easy to check for a
 specific structure-formation scenario whether it predicts
 an absolute upper limit to the
acceleration in halos of the order predicted by MOND. For example, the family
of halos produced in the simulations of \cite{nfw} do not seem to have a 
maximum acceleration, with higher-mass halos having higher accelerations
exceeding $\a0$ (Stacy McGaugh, Bob Sanders--private communications).
\par
2. The new limit is `mathematical'; i.e., it
 does not make further
assumptions on the physical nature of the galaxy. In contrast, the validity
of the previous limits rests on additional assumptions.
In the first example quasi-isothermality and
a nondegenerate-ideal-gas equation of state are assumed for the
 spherical system.
The limit then applies neither to normal stars, 
which are not isothermal, nor to
white dwarfs, whose equation of state is not that of an ideal gas.   
 These stars have, indeed, mean
 accelerations much higher than $\a0$. In the second example, instability
 is relied upon to cull out disks with high mean acceleration.
\par
3. The former two acceleration limits apply to the mean acceleration in the 
system, while the new limit applies to the `halo' acceleration at all radii. 


\end{document}